# On the Low SNR Capacity of Peak-Limited Non-Coherent Fading Channels with Memory

Amos Lapidoth    Ligong Wang

Swiss Federal Institute of Technology (ETH),

Zurich, Switzerland

e-mail: {lapidoth, wang}@isi.ee.ethz.ch

**Abstract**

The capacity of non-coherent stationary Gaussian fading channels with memory under a peak-power constraint is studied in the asymptotic weak-signal regime. It is assumed that the fading law is known to both transmitter and receiver but that neither is cognizant of the fading realization.

A connection is demonstrated between the asymptotic behavior of channel capacity in this regime and the asymptotic behavior of the prediction error incurred in predicting the fading process from very noisy observations of its past. This connection can be viewed as the low signal-to-noise ratio (SNR) analog of recent results by Lapidoth & Moser and by Lapidoth demonstrating connections between the high SNR capacity growth and the noiseless or almost-noiseless prediction error.

We distinguish between two families of fading laws: the "slowly forgetting" and the "quickly forgetting". For channels in the former category the low SNR capacity is achieved by IID inputs, whereas in the latter such inputs are typically sub-optimal. Instead, the asymptotic capacity can be approached by inputs with IID phase but block-constant magnitude.

## 1 Introduction

We present results on the low signal-to-noise ratio (SNR) asymptotic capacity of discrete-time peak-power limited single-antenna fading channels. The fading process is assumed to be a zero-mean unit-variance circularly-symmetric stationary complex Gaussian process with memory whose law—but not realization—is known to both the transmitter and the receiver. Input distributions that approach or achieve the asymptotic capacity are also studied.

Previous work on this channel [7], [5], [6] focused on the behavior of capacity at high SNR. For regular fading processes [7] demonstrated a connection between the high-SNR capacity growth and the prediction error in predicting the fading process from *noiseless* observations of its past, whereas for non-regular fading [6] demonstrated such a connection to the functional dependence on the observation noise variance of the prediction error in predicting the fading process from *noisy* observations of its past in the low observation noise regime.

Here we point to an analog connection between the low SNR asymptotic capacity and the prediction error in predicting the fading process from very noisy observations of its past. We show that, for most channels of interest, the limiting ratio of channel capacity to the square of the SNR is fully determined by the derivative at SNR = 0 of the prediction error in predicting the fading process based on observations of its past contaminated by IID Gaussian observation noise of variance 1/SNR.

Denoting this derivative by $(-\phi)$ we distinguish between two families of fading laws: the "slowly forgetting" where $\phi \geq 1/2$ and the "quickly forgetting" where $0 \leq \phi < 1/2$. For the former family IID inputs are asymptotically optimal, whereas for the latter IID inputs are typically (for $\phi > 0$) sub-optimal. To approach capacity on "quickly forgetting" fading channels we propose to use inputs of IID phase but block-constant magnitude.

When the fading spectral distribution function is discontinuous, i.e., when the fading has spectral lines, $\phi$ is infinite. In this case capacity at low SNR typically scales linearly with the SNR, with the limiting ratio of channel capacity to the SNR being the sum of the jumps in the spectral distribution function [12].

Previous work on the low SNR capacity of this channel dealt with the block fading model (which in the special case where the fading block size is one corresponds to memoryless fading for which $\phi = 0$) [2] and with the information rates achievable with fixed finite dimensional input distributions [9]. Our work relies heavily on both these papers.

It should be emphasized that this paper only deals with peak-power constraints. If those are replaced with average power constraints, capacity at low SNR typically scales linearly with the SNR; see [10], [8], [11] and references therein.

## 2 Channel Model and Main Results

### 2.1 Channel Model

We consider a discrete-time channel whose time-$k$ complex-valued output $Y_k \in \mathbb{C}$ is given by

$$Y_k = H_k x_k + Z_k \tag{1}$$

where $x_k \in \mathbb{C}$ is the complex-valued channel input at time $k$; the complex process $\{H_k\}$ models multiplicative noise; and the complex process $\{Z_k\}$ models additive noise.

We assume that the additive noise sequence $\{Z_k\}$ is a sequence of IID circularly symmetric complex-Gaussian random variables of mean zero and variance $\sigma^2$ where $\sigma^2 > 0$. Such a Gaussian distribution is denoted by $\mathcal{N}_\mathbb{C}(0, \sigma^2)$. The "fading process" $\{H_k\}$ is assumed to be a zero-mean, unit-variance, stationary, circularly symmetric, Gaussian process. We denote its autocorrelation function by $R(\cdot)$ and its spectral distribution function by $F(\cdot)$. Thus

$$R(m) \triangleq \mathsf{E}[H_{k+m}H_k^*] = \int_{-1/2}^{1/2} e^{\mathrm{i}2\pi m\lambda} \, \mathrm{d}F(\lambda), \qquad k,m \in \mathbb{Z} \tag{2}$$

and

$$R(0) = \mathsf{E}\big[|H_k|^2\big] = \int_{-1/2}^{1/2} \mathrm{d}F(\lambda) = 1. \tag{3}$$

We denote the derivative of the spectral distribution function by $F'(\cdot)$. If the spectral distribution function $F(\cdot)$ is absolutely continuous then the fading process has a spectral density. In this case the spectral density is given by $F'(\cdot)$ and will be denoted by $f(\cdot)$. Thus, if the fading has a spectral density $f(\cdot)$ then

$$R(m) = \int_{-1/2}^{1/2} e^{\mathrm{i}2\pi m\lambda} f(\lambda) \, \mathrm{d}\lambda, \qquad m \in \mathbb{Z} \tag{4}$$

and $f(\cdot)$ is a non-negative function satisfying

$$\int_{-1/2}^{1/2} f(\lambda) \, \mathrm{d}\lambda = 1. \tag{5}$$

We shall assume throughout that the processes $\{H_k\}$ and $\{Z_k\}$ are independent and that their joint law does not depend on the input sequence $\{x_k\}$.

The input constraint that we consider is the peak-power constraint according to which the time-$k$ input $x_k$ must satisfy

$$|x_k| \leq A \tag{6}$$

where $A$ is a positive real number and $A^2$ stands for the peak-power. We define the signal-to-noise ratio (SNR) as

$$\mathrm{SNR} \triangleq \frac{A^2}{\sigma^2}. \tag{7}$$

The capacity of this channel is given by

$$C(\mathrm{SNR}) = \lim_{n \to \infty} \frac{1}{n} \sup I(X_1, \ldots, X_n; Y_1, \ldots, Y_n) \tag{8}$$

where the supremum is over all joint distributions on $X_1, \ldots, X_n$ under which, with probability one,

$$|X_k| \leq A, \quad k = 1, \ldots, n. \tag{9}$$

## 2.2 Noisy Prediction

The least mean squared prediction error that one can attain when trying to predict $H_0$ based on its infinite past $(\ldots, H_{-2}, H_{-1})$ is given by [1]

$$\exp\left\{\int_{-1/2}^{1/2} \log F'(\lambda) \, \mathrm{d}\lambda\right\}. \tag{10}$$

More generally, if we try to predict $H_0$ based on $(\ldots, H_{-2}+W_{-2}, H_{-1}+W_{-1})$ where $\{W_k\}$ are IID $\mathcal{N}_{\mathbb{C}}(0, \delta^2)$ and independent of $\{H_k\}$ then the least mean squared prediction error $\epsilon^2_{\text{pred}}(\delta^2)$ is given by [6]

$$\epsilon^2_{\text{pred}}(\delta^2) = \exp\left\{\int_{-1/2}^{1/2} \log\left(F'(\lambda) + \delta^2\right) d\lambda\right\} - \delta^2. \tag{11}$$

The following lemma describes the asymptotics of the noisy prediction error from very noisy observations of the past:

**Lemma 1.** *If the fading process is of spectral density function $f(\lambda) \geq 0$ satisfying (5) and*

$$\int_{-1/2}^{1/2} f^2(\lambda) d\lambda < \infty \tag{12}$$

*then*

$$\phi \triangleq \lim_{\rho \downarrow 0} \frac{1 - \epsilon^2_{\text{pred}}(1/\rho)}{\rho} \tag{13}$$

*is defined and is given by*

$$\phi = \frac{1}{2} \int_{-1/2}^{1/2} f^2(\lambda) d\lambda - \frac{1}{2} \tag{14}$$

*which can also be expressed using Parseval's Theorem as*

$$\phi = \sum_{\nu=1}^{\infty} |R(\nu)|^2. \tag{15}$$

*Thus,*

$$\epsilon^2_{\text{pred}}(1/\rho) = 1 - \phi \cdot \rho + o(\rho) \tag{16}$$

*where the error term $o(\rho)$ is small enough so that $o(\rho)/\rho$ tends to zero as $\rho$ tends to zero.*

*Proof.* Using (11) we have

$$\frac{1 - \epsilon^2_{\text{pred}}(1/\rho)}{\rho} = \frac{1 - \left(\exp\left\{\int_{-1/2}^{1/2} \log(f(\lambda) + \rho^{-1}) d\lambda\right\} - \frac{1}{\rho}\right)}{\rho}$$

$$= \frac{1 - \left(\exp\left\{\int_{-1/2}^{1/2} [\log(1 + \rho f(\lambda)) + \log(1/\rho)] d\lambda\right\} - \frac{1}{\rho}\right)}{\rho}$$

$$= \frac{1}{\rho} + \frac{1}{\rho^2} - \frac{1}{\rho^2} \exp\left\{\int_{-1/2}^{1/2} \left(\rho f(\lambda) - \frac{1}{2}\rho^2 f^2(\lambda) + \Delta(\lambda, \rho)\right) d\lambda\right\}$$

$$= \frac{1}{\rho} + \frac{1}{\rho^2} - \frac{1}{\rho^2} \exp\left\{\rho - \frac{1}{2}\rho^2 \int_{-1/2}^{1/2} f^2(\lambda) d\lambda + \int_{-1/2}^{1/2} \Delta(\lambda, \rho)\right) d\lambda\right\}$$

where
$$\Delta(\lambda, \rho) \triangleq \log(1 + \rho f(\lambda)) - \left(\rho f(\lambda) - \frac{1}{2}\rho^2 f^2(\lambda)\right).$$

The result now follows by a second order Taylor expansion of the exponential function and from the fact that

$$\lim_{\rho \downarrow 0} \int_{-1/2}^{1/2} \frac{\Delta(\lambda, \rho)}{\rho^2} \, d\lambda = 0. \tag{17}$$

This latter fact can be proved using the Monotone Convergence Theorem (MCT). Indeed, for any fixed $\lambda \in [-1/2, 1/2]$ we have by Taylor's expansion applied to the function $\xi \mapsto \log(1 + \xi)$ that the integrand converges to zero as $\rho \downarrow 0$. Moreover, it can be verified that for such fixed $\lambda$ the integrand is monotonically decreasing to zero as $\rho \downarrow 0$. Finally, for $\rho = 1$ the integrand is an integrable function. □

## 2.3 Main Results

The asymptotic behavior of the capacity at low SNR for fading processes having a spectral density function is given in the following theorem.

**Theorem 1.** *If the stationary zero-mean circularly-symmetric complex Gaussian fading process has a spectral distribution function $f(\lambda) \geq 0$ satisfying (5) and (12) then*

$$\lim_{\text{SNR} \downarrow 0} \frac{C(\text{SNR})}{\text{SNR}^2} = \begin{cases} \frac{1}{8}(2\phi + 1)^2 & \text{if } \phi < \frac{1}{2} \\ \phi & \text{if } \phi \geq \frac{1}{2} \end{cases} \tag{18}$$

*where $\phi$ is given in (14) or (15).*

We refer to fading processes for which $\phi \geq 1/2$ as "slowly forgetting" and to those with $0 \leq \phi < 1/2$ as "quickly forgetting". In the former case when $\phi \geq 1/2$ the asymptotic capacity is achieved by IID inputs taking the values $\pm A$ with equal probability. For the latter case when $0 < \phi < 1/2$ the asymptotic capacity is typically not achieved by IID inputs [12]. In such cases, information rates that at low SNR get arbitrarily close to the capacity can be achieved by inputs of IID phase but block-constant magnitude.

When the spectral distribution function is discontinuous, the fading process has spectral lines and does not have a spectral density function. Nevertheless, being a monotonic function, the spectral distribution function is differentiable except on a set of measure zero. In this case capacity typically increases linearly with the SNR in the low SNR regime with the limiting ratio of capacity to SNR being the sum of the jumps in the spectral distribution function $F(\cdot)$ [12].

Theorem 1 will be proved in Section 3 by exhibiting upper and lower bounds on channel capacity that at low SNR coincide.

# 3 Proof of Theorem 1

In this section we prove Theorem 1 by deriving an upper bound and a lower bound on channel capacity that coincide at low SNR. We begin with the upper bound.

## 3.1 Proof of the Upper Bound

Our first steps towards an upper bound on channel capacity are similar to those of [6]. We begin with the chain rule:

$$
\begin{aligned}
I(X_1^n; Y_1^n) &= \sum_{k=1}^{n} I\left(X_1^n; Y_k \big| Y_1^{k-1}\right) \\
&= \sum_{k=1}^{n} I\left(X_1^n, Y_1^{k-1}; Y_k\right) - I\left(Y_k; Y_1^{k-1}\right) \\
&\leq \sum_{k=1}^{n} I\left(X_1^n, Y_1^{k-1}; Y_k\right) \\
&= \sum_{k=1}^{n} I\left(X_1^k, Y_1^{k-1}; Y_k\right) \\
&= \sum_{k=1}^{n} I(X_k; Y_k) + I\left(X_1^{k-1}, Y_1^{k-1}; Y_k \big| X_k\right).
\end{aligned}
\tag{19}
$$

Consequently we have by (8)

$$
C(\mathrm{SNR}) \leq \sup_{p_{X_0}} \left\{ I(X_0; Y_0) + I\left(X_{-\infty}^{-1}, Y_{-\infty}^{-1}; Y_0 \big| X_0\right) \right\}
\tag{20}
$$

where the term $I(X_0; Y_0)$ is the mutual information corresponding to the input distribution $p_{X_0}$ for the memoryless Rayleigh fading channel. By [2]

$$
I(X_0; Y_0) = \frac{\mathsf{E}\left[|X_0|^4\right] - \left(\mathsf{E}\left[|X_0|^2\right]\right)^2}{2\sigma^4} + o\left(\mathrm{SNR}^2\right),
\tag{21}
$$

where $o\left(\mathrm{SNR}^2\right)/\mathrm{SNR}^2 \to 0$ as $\mathrm{SNR} \to 0$.

We now study the second term on the right hand side of (20). As in [6] we introduce random variables $\{W_k\}$ that are IID $\mathcal{N}_{\mathbb{C}}(0, 1/\mathrm{SNR})$ and independent of $\{H_k\}$ to model the observation noise. Denoting by $\mathsf{E}[\cdot]$ the mathematical expectation we have

$$
\begin{aligned}
I\left(X_{-\infty}^{-1}, Y_{-\infty}^{-1}; Y_0 \big| X_0\right) &= h\left(Y_0 \big| X_0\right) - h\left(Y_0 \big| X_0, X_{-\infty}^{-1}, Y_{-\infty}^{-1}\right) \\
&\leq h\left(Y_0 \big| X_0\right) - h\left(Y_0 \big| X_0, \{H_\nu + W_\nu\}_{\nu=-\infty}^{-1}\right) \\
&= \mathsf{E}\left[\log \frac{\sigma^2 + |X_0|^2}{\sigma^2 + |X_0|^2 \cdot \epsilon_{\mathrm{pred}}^2(1/\mathrm{SNR})}\right]
\end{aligned}
$$

$$\leq \log \frac{\sigma^2 + \mathsf{E}\big[|X_0|^2\big]}{\sigma^2 + \mathsf{E}[|X_0|^2] \cdot \epsilon_{\text{pred}}^2(1/\text{SNR})}$$

$$= \log \frac{\sigma^2 + \mathsf{E}\big[|X_0|^2\big]}{\sigma^2 + \mathsf{E}[|X_0|^2] \cdot (1 - \phi \cdot \text{SNR} + o(\text{SNR}))}$$

$$\leq \phi \cdot \text{SNR} \cdot \frac{\mathsf{E}\big[|X_0|^2\big]}{\sigma^2} + o\left(\text{SNR}^2\right). \tag{22}$$

Here the first inequality follows because, given the present input, all the information about the present output that is contained in the past inputs and outputs is maximized when all the past inputs have maximum magnitude; the subsequent equality follows from the explicit expression for the differential entropy of circularly-symmetric Gaussians and because we assumed that the fading process is of unit-variance; the subsequent inequality follows from Jensen's inequality, because $(\sigma^2 + r)/(\sigma^2 + \epsilon^2 r)$ is concave in $r \in [0, \infty)$ when $\epsilon^2 \leq 1$ and therefore so is its logarithm; the subsequent equality by Lemma 1; and the final step by expressing the logarithm of the ratio as the difference of two logarithms and by expanding each of the logarithms into its Taylor series expansion keeping in mind that by the peak-constraint $|X_0|^2/\sigma^2 \leq \text{SNR}$.

By (20), (21), and (22) we obtain

$$C \leq \sup_{p_{X_0}} \left\{ \frac{\mathsf{E}\big[|X_0|^4\big] - \big(\mathsf{E}\big[|X_0|^2\big]\big)^2}{2\sigma^4} + \phi \cdot \text{SNR} \cdot \frac{\mathsf{E}\big[|X_0|^2\big]}{\sigma^2} \right\} + o\left(\text{SNR}^2\right) \tag{23}$$

$$\leq \sup_{p_{X_0}} \left\{ \frac{A^2 \mathsf{E}\big[|X_0|^2\big] - \big(\mathsf{E}\big[|X_0|^2\big]\big)^2}{2\sigma^4} + \phi \cdot \text{SNR} \cdot \frac{\mathsf{E}\big[|X_0|^2\big]}{\sigma^2} \right\} + o\left(\text{SNR}^2\right) \tag{24}$$

where the second inequality follows by noting that, by the peak-constraint (6), we have $0 \leq |X_0| \leq A$ so that $\mathsf{E}\big[|X_0|^4\big] \leq A^2 \mathsf{E}\big[|X_0|^2\big]$. Also by (6) we have that

$$0 \leq \frac{\mathsf{E}\big[|X_0|^2\big]}{A^2} \leq 1 \tag{25}$$

so that if we introduce

$$\alpha \triangleq \mathsf{E}\big[|X_0|^2\big] / A^2 \tag{26}$$

then $0 \leq \alpha \leq 1$. Consequently we have from (24) that

$$C \leq \sup_{0 \leq \alpha \leq 1} \left\{ \frac{\alpha - \alpha^2}{2} + \phi \cdot \alpha \right\} \cdot \text{SNR}^2 + o\left(\text{SNR}^2\right). \tag{27}$$

When $\phi < 1/2$, the supremum on the RHS of (27) is achieved inside the interval $[0, 1]$ by $\alpha^* = \phi + \frac{1}{2}$ to yield

$$C \leq \frac{(2\phi + 1)^2}{8} \text{SNR}^2 + o\left(\text{SNR}^2\right), \qquad \phi < \frac{1}{2} \tag{28}$$

and when $\phi \geq 1/2$ the supremum is achieved at $\alpha^* = 1$ to yield

$$C \leq \phi \cdot \text{SNR}^2 + o\left(\text{SNR}^2\right), \qquad \phi \geq \frac{1}{2}. \tag{29}$$

In both inequalities $o\left(\text{SNR}^2\right)/\text{SNR}^2 \to 0$ as $\text{SNR} \to 0$.

## 3.2 Proof of the Lower Bound

We shall next exhibit a lower bound on channel capacity that at low SNR coincides with the upper bound (27). More specifically, for any $0 \leq \alpha \leq 1$ of our choice and for any integer $b \geq 1$ we shall propose for every peak-amplitude $A > 0$ a distribution on $\{X_k\}_{k=1}^\infty$ satisfying the peak-constraint (6) such that

$$\lim_{b \to \infty} \lim_{\text{SNR} \downarrow 0} \frac{\lim_{n \to \infty} \frac{1}{n} I(X_1^n; Y_1^n)}{\text{SNR}^2} \geq \frac{\alpha - \alpha^2}{2} + \phi \cdot \alpha, \quad 0 \leq \alpha \leq 1. \tag{30}$$

The proposed distribution on $\{X_k\}$ can be described as follows:

$$X_k = U_{\lfloor k/b \rfloor} \cdot D_k, \quad k = 1, 2, \ldots \tag{31}$$

where $\lfloor \xi \rfloor$ denotes the largest integer not exceeding $\xi$; $\{D_k\}_{k=1}^\infty$ are IID taking the values $\pm 1$ equi-probably; $\{U_\nu\}_{\nu=0}^\infty$ are IID taking the value $A$ with probability $\alpha$ and the value 0 with probability $(1-\alpha)$; and the sequences $\{D_k\}_{k=1}^\infty$ and $\{U_\nu\}_{\nu=0}^\infty$ are independent.

We shall next show that for this distribution

$$\lim_{n \to \infty} \frac{1}{n} I(X_1^n; Y_1^n) \geq \frac{1}{b} I(X_1, \ldots, X_b; Y_1, \ldots, Y_b). \tag{32}$$

This is a simple consequence of the fact that $\{X^{(\nu)}\}$ are IID where $X^{(\nu)}$ is the $\nu$-th tuple

$$X^{(\nu)} = (X_{\nu b+1}, \ldots, X_{\nu b+b}), \quad \nu = 0, 1, \ldots$$

and where we analogously define

$$Y^{(\nu)} = (Y_{\nu b+1}, \ldots, Y_{\nu b+b}), \quad \nu = 0, 1, \ldots$$

Indeed, defining $m = \lfloor n/b \rfloor$ we have

$$\frac{1}{n} I(X_1^n; Y_1^n) \geq \frac{1}{n} I\left(\{X^{(\nu)}\}_{\nu=0}^{m-1}; Y_1^n\right)$$

$$= \frac{1}{n} \sum_{\nu=0}^{m-1} I\left(X^{(\nu)}; Y_1^n \Big| \{X^{(\eta)}\}_{\eta=0}^{\nu-1}\right)$$

$$= \frac{1}{n} \sum_{\nu=0}^{m-1} I\left(X^{(\nu)}; Y_1^n, \{X^{(\eta)}\}_{\eta=0}^{\nu-1}\right)$$

$$\geq \frac{1}{n} \sum_{\nu=0}^{m-1} I\left(X^{(\nu)}; Y^{(\nu)}\right)$$

$$= \frac{m}{n} I\left(X^{(0)}; Y^{(0)}\right)$$

$$= \frac{\lfloor n/b \rfloor}{n} I(X_1, \ldots, X_b; Y_1, \ldots, Y_b)$$

from which (32) follows upon letting $n$ tend to infinity.

Having proved (32) we now have that for our proposed input distribution

$$\lim_{\text{SNR}\downarrow 0} \frac{\lim_{n\to\infty} \frac{1}{n}I\left(X_1^n;Y_1^n\right)}{\text{SNR}^2} \geq \frac{1}{b} \lim_{\text{SNR}\downarrow 0} \frac{I(X_1,\ldots,X_b;Y_1,\ldots,Y_b)}{\text{SNR}^2}. \tag{33}$$

We next use Corollary 1 of [9] to study the RHS of the above noting that for the proposed input distribution (31)

$$\mathsf{E}\left[X_k X_j^*\right] = \begin{cases} \alpha \cdot A^2 & \text{if } k=j \\ 0 & \text{otherwise} \end{cases} \tag{34}$$

and

$$\mathsf{E}\left[|X_k|^2 |X_j|^2\right] = \alpha \cdot A^4, \quad j,k \in \{1,\ldots,b\}. \tag{35}$$

Let $\mathbf{X}$ denote the vector $(X_1, X_2, \ldots, X_b)^\mathsf{T}$, and let $\mathbb{H}$ denote the diagonal matrix with diagonal entries $H_1, H_2, \ldots, H_b$, then

$$I\left(X_1^b; Y_1^b\right) = \frac{1}{2\sigma^4}\text{tr}\left\{\mathsf{E}\left[\left(\mathsf{E}\left[\mathbb{H}\mathbf{X}\mathbf{X}^\dagger \mathbb{H}^\dagger | \mathbf{X}\right]\right)^2\right] - \left(\mathsf{E}\left[\mathbb{H}\mathsf{E}\left[\mathbf{X}\mathbf{X}^\dagger\right]\mathbb{H}^\dagger\right]\right)^2\right\} + o\left(\text{SNR}^2\right)$$

$$= \frac{1}{2\sigma^4}\text{tr}\left\{\mathsf{E}\left[\begin{pmatrix} R(0)|X_1|^2 & \cdots & R(1-b)X_1 X_b^* \\ \vdots & \ddots & \vdots \\ R(b-1)X_b X_1^* & \cdots & R(0)|X_b|^2 \end{pmatrix}^2\right]\right.$$

$$\left. - \left(\mathsf{E}\left[\mathbb{H}\begin{pmatrix} \mathsf{E}\left[|X_1|^2\right] & & \\ & \ddots & \\ & & \mathsf{E}\left[|X_b|^2\right] \end{pmatrix} \mathbb{H}^\dagger\right]\right)^2\right\} + o\left(\text{SNR}^2\right)$$

$$= \frac{1}{2\sigma^4}\left(\sum_{i=1}^{b}\sum_{j=1}^{b}|R(i-j)|^2 \mathsf{E}\left[|X_i|^2|X_j|^2\right] \right.$$

$$\left. - \sum_{i=1}^{b}|R(0)|^2\left(\mathsf{E}\left[|X_i|^2\right]\right)^2\right) + o\left(\text{SNR}^2\right)$$

$$= \frac{1}{2\sigma^4}\left(\alpha \cdot A^4 \sum_{i=1}^{b}\sum_{j=1}^{b}|R(i-j)|^2 - b\cdot\alpha^2\cdot A^4\right) + o\left(\text{SNR}^2\right)$$

$$= \frac{1}{2}\text{SNR}^2\left(b\cdot(\alpha-\alpha^2) + \alpha\sum_{i=1}^{b}\sum_{\substack{1\leq j\leq b \\ j\neq i}}|R(i-j)|^2\right) + o\left(\text{SNR}^2\right)$$

$$= \frac{1}{2}\text{SNR}^2\left(b\cdot(\alpha-\alpha^2) + \alpha S(b)\right) + o\left(\text{SNR}^2\right) \tag{36}$$

where in the last equality we defined

$$S(b) \triangleq \sum_{i=1}^{b} \sum_{\substack{1\leq j\leq b \\ j\neq i}} |R(i-j)|^2. \tag{37}$$

Dividing (36) by $b$ we obtain

$$\frac{1}{b}I\left(X_1^b;Y_1^b\right) = \frac{1}{2}\left(\alpha - \alpha^2 + \alpha\frac{S(b)}{b}\right)\text{SNR}^2 + o\left(\text{SNR}^2\right). \tag{38}$$

Using (38) and (33) the proof of (30) is completed by noting that

$$\lim_{b\to\infty}\frac{S(b)}{b} = 2\phi. \tag{39}$$

which follows from a direct calculation showing that

$$S(b+1) - S(b) = 2\sum_{\eta=1}^{b}|R(\eta)|^2 \tag{40}$$

$$\to 2\phi \qquad (\text{as } b \to \infty) \tag{41}$$

and a Cesáro argument.

**Remark 1.** *When $\phi \geq 1/2$, the optimal choice for $\alpha$ is 1. This corresponds to the sequence $\{U_\nu\}$ in (31) being deterministically equal to A, thus resulting in our proposed input sequence $\{X_k\}$ being IID. It is thus seen that for $\phi \geq 1/2$ the asymptotic capacity at low SNR is achieved by IID inputs. For $\phi < 1/2$ our proposed input distribution is not IID. Indeed, for $0 < \phi < 1/2$ IID inputs typically do not achieve the low SNR capacity [12].*

## Acknowledgment

The authors gratefully acknowledge stimulating discussions with T. Koch.